\documentclass[pra,amsmath,amssymb,twocolumn,superscriptaddress]{revtex4-1}

\usepackage{amsthm,amssymb,amsmath} %
\usepackage{dcolumn}
\usepackage{bm}
\usepackage{graphicx}
\usepackage{color}
\usepackage{subfig,array}
\usepackage[breaklinks]{hyperref}
\usepackage[singlelinecheck=off,justification=raggedright]{caption} %

\begin{document}

\newtheorem{corollary}{Corollary}
\newtheorem{definition}{Definition}
\newtheorem{example}{Example}
\newtheorem{lemma}{Lemma}
\newtheorem{proposition}{Proposition}
\newtheorem{theorem}{Theorem}
\newtheorem{fact}{Fact}
\newtheorem{property}{Property}
\newtheorem{observation}{Observation}
\newcommand{\bra}[1]{\langle #1|}
\newcommand{\ket}[1]{|#1\rangle}
\newcommand{\braket}[3]{\langle #1|#2|#3\rangle}
\newcommand{\ip}[2]{\langle #1|#2\rangle}
\newcommand{\op}[2]{|#1\rangle \langle #2|}

\newcommand{\tr}{{\rm tr}}
\newcommand {\E } {{\mathcal{E}}}
\newcommand {\F } {{\mathcal{F}}}
\newcommand {\diag } {{\rm diag}}
\newcommand{\comment}[1]{{\color{red} #1}}

\title{Dichotomy of entanglement depth for symmetric states}

\author{Ji-Yao Chen}%
\affiliation{State Key Laboratory of Low Dimensional Quantum Physics, Department of Physics, Tsinghua University, Beijing, China}%
\affiliation{Perimeter Institute for Theoretical Physics, Waterloo, Ontario, Canada}

\author{Zhengfeng Ji}
\affiliation{Centre for Quantum Computation \& Intelligent
Systems,
Faculty of Engineering and Information Technology, University of
Technology, Sydney, NSW 2007, Australia}%

\author{Nengkun Yu}
\affiliation{Institute for Quantum Computing, University of Waterloo,
  Waterloo, Ontario, Canada}%
\affiliation{Department of Mathematics \& Statistics, University of
  Guelph, Guelph, Ontario, Canada}%
\affiliation{Centre for Quantum Computation \& Intelligent
Systems,
Faculty of Engineering and Information Technology, University of
Technology, Sydney, NSW 2007, Australia}

\author{Bei Zeng}
\affiliation{Institute for Quantum Computing, University of Waterloo,
  Waterloo, Ontario, Canada}%
\affiliation{Department of Mathematics \& Statistics, University of
  Guelph, Guelph, Ontario, Canada}%
\affiliation{Canadian Institute for Advanced Research, Toronto,
  Ontario, Canada}%

\begin{abstract}
  Entanglement depth characterizes the minimal number of particles in
  a system that are mutually entangled. For symmetric states, 
  there is a dichotomy for entanglement depth: an $N$-particle
  symmetric state is either fully separable, or fully entangled ---
  the entanglement depth is either $1$ or $N$. We show that
  this dichotomy property for entangled symmetric states
  is even stable under non-symmetric noise. 
  We propose an experimentally
  accessible method to detect entanglement depth in atomic ensembles
  based on a bound on the particle number population of Dicke states,
  and demonstrate that the entanglement depth of some Dicke states,
  for example the twin Fock state, is very stable even under a large
  arbitrary noise. Our observation can be applied to atomic
  Bose-Einstein condensates to infer that these systems can be highly
  entangled with the entanglement depth that is of the order of the
  system size (i.e. several thousands of atoms).
\end{abstract}

\maketitle

\section{Introduction}

Entanglement, a kind of quantum correlations that is stronger than any possible classical correlation, is one of the central mystery of quantum mechanics~\cite{horodecki2009quantum,guhne2009entanglement,eckert2002quantum}. It is now recognized as a resource for secure communication, quantum information processing, and precision measurements~\cite{giovannetti2006quantum,ji2008parameter}. A many-body quantum state may be highly entangled, and the degree of such entanglement can be characterized by different kinds of entanglement measures, such as $n$-tangle, geometric entanglement and tensor rank~\cite{horodecki2009quantum}.

Among these many-body entanglement measures, a natural measure is the entanglement depth~\cite{sorensen2001entanglement}, which characterizes the minimal number of particles that are entangled. More precisely, an $N$-particle state $\rho_N$ has entanglement depth $\mathcal{D}$, if for all decomposition of $\rho_N=\sum_i p_i\ket{\psi^i_N}\bra{\psi^i_N}$, there exists one pure state $\ket{\psi_N^i}$ that has at least $\mathcal{D}$-particle entanglement. That is, $\rho_N$ has `genuine' $\mathcal{D}$-particle entanglement. We use $\mathcal{D}(\rho_N)$ to denote the entanglement depth of $\rho_N$. Obviously in general, all $1\leq \mathcal{D}(\rho_N)\leq N$ are possible for $\rho_N$.

Entanglement depth is widely used as a measure to characterize entanglement properties, both theoretically
and experimentally. It is in particular applicable for characterizing entanglement properties
in the cold atomic systems, where the particle number $N$ of systems could be up to order of several thousands~\cite{sorensen2001entanglement,lucke2014detecting,toth2014quantum,
duan2011entanglement,korbicz2006generalized,ma2011quantum,vitagliano2014spin,korbicz2005spin,mcconnell2015entanglement}, and entanglement depth as large as $\sim 3000$ is reported in~\cite{mcconnell2015entanglement}.

An $N$-particle wavefunction $\ket{\psi_N}$ is symmetric if it is invariant under permutation of any two particles. An $N$-particle state $\rho_N$ is symmetric if it is a mixture of symmetric wavefunctions. Entanglement for symmetric states have been extensively studied~\cite{stockton2003characterizing,ichikawa2008exchange,wei2010exchange,toth2009entanglement,nengkun2015}. Experimentally, there exist systems with symmetric wave functions. For example, for an atomic Bose-Einstein condensate (BEC), in case the single-mode approximation is valid, the internal states of the atoms, under ideal situation, can be thought as symmetric states of $N$-particles, where each particle is of finite dimension~\cite{leggett2001bose}.

For symmetric states, there is a dichotomy for entanglement depth: an $N$-particle symmetric state is either fully separable, or fully entangled -- i.e. the entanglement depth is either $1$ or $N$~\cite{ichikawa2008exchange,wei2010exchange}. This indicates that, in the ideal case, if the state $\rho_N$ of the system is fully symmetric, $\mathcal{D}(\rho_N)\geq 2$ in fact means $\mathcal{D}(\rho_N)=N$. In other words, under ideal situation, an entangled symmetric state always has entanglement depth of the system size $N$ (i.e. the number of particles in the system). For example, in a typical BEC system, the system size $N$ can be the order of several thousands. However, in practice, it is very hard to maintain a fully symmetric state,
so it is important to understand the stability of entanglement depth.

In this work, we show that the entanglement depth is stable
even under non-symmetric noise. The range of the noise that each
symmetric state may tolerate depends on the state. However, there
does exist symmetric states whose entanglement depth is highly stable.
One such state is the twin Fock state, which is very
stable even under strong arbitrary noise. 
We then propose an experimentally accessible method to
detect entanglement depth based on a bound of the particle number
population of Dicke states, a detection technique that is widely
available in various systems, such as the cold atomic systems.

We compare our results with other methods for detecting entanglement depth,
such as spin squeezing parameters~\cite{sorensen2001entanglement,lucke2014detecting,gross2012spin,toth2014quantum,
duan2011entanglement,korbicz2006generalized,ma2011quantum,vitagliano2014spin,korbicz2005spin} and quantum Fisher information~\cite{gross2012spin,toth2014quantum}. In a sharp contrast,
for symmetric states, despite the dichotomy of entanglement
depth, spin squeezing is very unlikely. This is because in order to detect entanglement of the state $\rho_N$ by the spin-squeezing criteria, the two-particle reduced density matrix of $\rho_N$ must be entangled, which is very unlikely
due to the quantum finite de Finetti's theorem~\cite{stormer1969symmetric,hudson1976locally,doherty02a,Renner2007,Christandl2007,harrow2013church}. Compared to spin squeezing parameters, quantum Fisher information is more useful to detect entanglement depth for symmetric state~\cite{oszmaniec2016random}, however in many cases it still cannot reveal faithfully that an entangled symmetric state always have entanglement depth $N$.

Our results clarify an important fact regarding the entanglement properties of symmetric states, that is, an entangled symmetric state is always highly entangled, an intrinsic property arising from permutation symmetry. These kinds of entanglement are also very stable, even stable under large non-symmetric noise for certain states, and can also be effectively detected via simple method through measuring particle number population, a detection technique that is widely available. This makes certain physical systems, such as BECs, ideal systems for generating and detecting many-particle entanglement given that they are `born with symmetry'.  Even very weak interactions in these systems can create highly entangled states~\cite{mcconnell2015entanglement}, whose scale of entanglement is not currently reachable by any other technology.  

\section{Stability of entanglement depth for symmetric states} 

Consider an $N$-particle system $\mathcal{H}_N=\mathbb{C}^{d_1}\otimes \mathbb{C}^{d_2}\otimes\cdots \mathbb{C}^{d_N}$. An $N$-particle pure state $\ket{\psi_N}$ is called genuine entangled if it is not a product state of any bipartition. $\ket{\psi_N}$ is called $k$-product (separable) if it can be written as
\begin{equation}
\ket{\psi_N}=\ket{\psi^{(1)}}\otimes\ket{\psi^{(2)}}\cdots\otimes\ket{\psi^{(k)}},
\end{equation}
where decomposition corresponds to a partition of the $N$ particles, $\ket{\psi^{(i)}}$ is a genuine entangled state in $\mathcal{H}_{S_i}=\otimes_{j\in S_i} \mathbb{C}^{d_j}$ with $\bigcup S_i=\{1,2,\cdots,N\}$ and $S_i\bigcap S_k=\varnothing$ for $i\neq k$. The entanglement depth of $\ket{\psi_N}$, $\mathcal{D}(\psi_N)$, is defined as the largest cardinality of $S_i$.

An $N$-particle density matrix $\rho_N$ is called
$k$-separable if it can be written as some convex combination of $k$-separable pure states. The entanglement depth of $\rho_N$ is defined as following \cite{sorensen2001entanglement},
\begin{equation}
\mathcal{D}(\rho_N)=\min_{\rho_N=\sum p_i \op{\psi_N^i}{\psi_N^i}}\max_i~~ \mathcal{D}(\psi_N^i),
\end{equation}
where each $\ket{\psi_N^i}$ is an $N$-particle pure state and $\mathcal{D}(\psi_N^i)$ is the entanglement depth of $\ket{\psi_N^i}\bra{\psi_N^i}$.

An $N$-particle density matrix $\rho_N$ has entanglement depth at least $N/k$ if
it is not $k$-separable. That is, the entanglement depth $1$ corresponds to a fully separable state ($N$-separable), and the entanglement depth $N$ corresponds to a genuine entangled state (i.e. not bi-separable). In this sense, entanglement depth characterizes the minimal number of particles in the system that are not separable~\cite{sorensen2001entanglement}.

Symmetric states are a kind of many-body states that attract a lot of attention. An $N$-particle pure state $\ket{\psi_N}$ is symmetric if it is invariant under permutation of particles. That is, $\ket{\psi_N}$ lives inside the symmetric subspace $\mathcal{H}^+_N$ of $\mathcal{H}_N=(\mathbb{C}^d)^{\otimes N}$. A symmetric $N$-particle density matrix $\rho_N$ is a mixture of symmetric pure states, i.e., $\rho_N$ is supported on the symmetric subspace $\mathcal{H}^+_N$.
A symmetric $N$-particle state is a density matrix that is supported on the symmetric subspace $\mathcal{H}^+_N$.

The following fact is known for symmetric states~\cite{ichikawa2008exchange,wei2010exchange}: the entanglement depth of an $N$-particle symmetric state is either $1$ or $N$.

For completeness, we include a proof of this fact. We first look at the pure state case. Suppose $\ket{\psi}$ is a symmetric state with entanglement depth less than $N$, then it can be written as $\ket{\alpha}\otimes\ket{\beta}$ for some bipartition with some $\ket{\alpha}$ and $\ket{\beta}$. Employing the symmetry of $\ket{\psi}$, we know that for any bipartition, $\ket{\psi}$ is in product form. That is, $\ket{\psi}=\ket{\gamma}^{\otimes N}$ for some single-particle state $\ket{\gamma}$. This indicates that the entanglement depth is $1$.

For general mixed state, we suppose $N$-particle symmetric state $\rho$ has entanglement depth less than $N$. Then, it can be written as convex combination of some bipartite product pure states. The bi-partitions can be different.  More precisely, $\rho_N=\sum p_i \op{\psi_N^i}{\psi_N^i}$ with $p_i>0$ where $\ket{\psi_N^i}$s are all bi-separable symmetric state. According to the pure state case discussed above, we know that $\ket{\psi_N^i}=\ket{\gamma_i}^{\otimes N}$ for some single-particle state $\ket{\gamma_i}$. Then we confirm that if the entanglement depth of an $N$-particle symmetric state is less than $N$, its entanglement depth is $1$.

That is, any entangled many-body symmetric state is born to be `highly entangled'. However, in practice, if any non-symmetric perturbation destroys this large entanglement depth, then one cannot really observe this large entanglement in physical systems due to uncontrollable environmental noise.
In the next, we show that, fortunately, the entanglement depth is stable under general (not necessarily symmetric) perturbation.

\begin{observation}
Entanglement depth is stable against small perturbations. In particular, for an $N$-particle state $\rho_N$ with $\mathcal{D}(\rho_N)=N$, the entanglement depth is stable against small perturbation.
\end{observation}


To see the validity of this observation, 
notice that the set of bi-separable states is convex compact in the set of all states by applying the Caratheodory's theorem about convex hull~\cite{Eckhoff1993}.
Therefore, the set of genuinely entangled states, the complementary set of bi-separable states, is open.  This indicates that for any $\rho$ with entanglement depth $N$, there exists $\epsilon_{\rho}>0$ such that the open ball with center $\rho$ and radius $\epsilon_{\rho}$, $B:=\{\sigma\lvert |\sigma-\rho|<\epsilon_{\rho}\}$ satisfies that the entanglement depth of $\sigma\in B$ is $N$. In other words, genuine entanglement is robust against small perturbation.

One can use similar argument to obtain that: In multipartite system consisting of $N$ particles, the set of states with entanglement depth no more than $r\leq N$ forms an convex compact (closed) set. Thus, its complementary set, the set of states with entanglement depth greater than $r$ forms an open set. This indicates that for any $\rho$ with entanglement depth $r+1$, there exists $\epsilon_{\rho}>0$ such that the open ball with center $\rho$ and radius $\epsilon_{\rho}$, $B:=\{\sigma\lvert |\sigma-\rho|<\epsilon_{\rho}\}$ satisfies that the entanglement depth of $\sigma\in B$ is at least $r+1$.

This then shows that the entanglement depth of any $N$-particle entangled
symmetric state $\rho_N$ is non-decreasing. The stability region
depends on each state. For example, an entangled symmetric state that
is very close to a separable state ($e.g.$ $\ket{0}^{\otimes N}$) has
relatively small stability region. However, some other state, for
example the twin Fock state as we will discuss later, is relatively
very stable. In other words, experimentally prepared twin Fock state
has entanglement depth that is very stable against noise.

\section{Detecting entanglement depth}

The entanglement depth $\mathcal{D}(\rho_N)$, by itself, is not a physical observable. It is therefore, important to propose physical observables that can be used to infer entanglement depth of the system.  In the following, we discuss a general method to detect entanglement depth, by measuring particle number populations of Dicke states. We consider ensembles of particles with $d$ dimensional single particle Hilbert space, which are equivalently spin $S=(d-1)/2$ particles.

For simplicity we consider $d=2$, and similar methods can naturally extend to $d>2$. The single particle space is then of dimension $2$ and we denote the basis by $\ket{0},\ket{1}$.
In this setting, the symmetric subspace of an $N$-particle system is of dimension $N+1$ and is spanned by the Dicke states $\ket{D_{N,r}}$ with $0\leq r\leq N$, which are the normalized symmetric states with $r$ excitations (i.e. $r$ number of 1s).
It is obvious that all $\ket{D_{N,r}}$ are entangled except $\ket{D_{N,0}}$ and $\ket{D_{N,N}}$.

In general, an $N$-particle Hilbert space has dimension $d^N$ ($2^N$ for $d=2$).
If we experimentally prepare a target Dicke state $\ket{D_{N,r}}$, and measure the number $N_r$ of
particle population on $\ket{D_{N,r}}$. Since the total number of particles is $N$, there is
then $n_r=\frac{N_r}{N}$ fractions of particles that are populated on $\ket{D_{N,r}}$. We then further
treat the rest $N-N_r$ particles of the system as `noise' since they do not populate on the desired state $\ket{D_{N,r}}$. We can then model the real state of the system in the form (denote
$\rho_{N,r}=\ket{D_{N,r}}\bra{D_{N,r}}$)
\begin{equation}
\tilde{\rho}=n_r\rho_{N,r}+(1-n_r)\rho_{\text{noise}},
\end{equation}
where $n_r=\langle D_{N,r}|\tilde{\rho}\ket{D_{N,r}}$.

Our goal is to estimate the
entanglement depth of $\tilde{\rho}$. We start from the most general case that $\rho_{\text{noise}}$ is an arbitrary noise, which might not be a positive operator. Under this assumption,  we
find a threshold value $p_{N,r}$, such that any $n_r>p_{N,r}$ ensures $\tilde{\rho}$ has entanglement depth $N$ (i.e. fully entangled).
The threshold value can be estimate by the following observation.
\begin{observation}
\label{obv:3}
For any bi-separable $N$-particle state $\rho_N$, we have
\begin{equation}
\langle D_{N,r}|\rho_N\ket{D_{N,r}}\leq p_{N,r}.
\end{equation}
The value of $p_{N,r}$ is given by
\begin{equation}
p_{N,r}:=\max\{{\frac{C_{m_0}^{j}C_{m_1}^{r-j}}{C_N^{r}}}\},
\end{equation}
where $m_0+m_1=N, m_0,m_1\geq 1,j\geq 0$. As a direct consequence, any $N$-particle state $\sigma_N$ satisfying $\langle D_{N,r}|\sigma_N\ket{D_{N,r}}> p_{N,r}$ is not bi-separable, thus, has entanglement depth $N$.
\end{observation}

\begin{figure}[hbpt]
\centering
\subfloat[]{
\includegraphics[width=40mm,height=35mm]{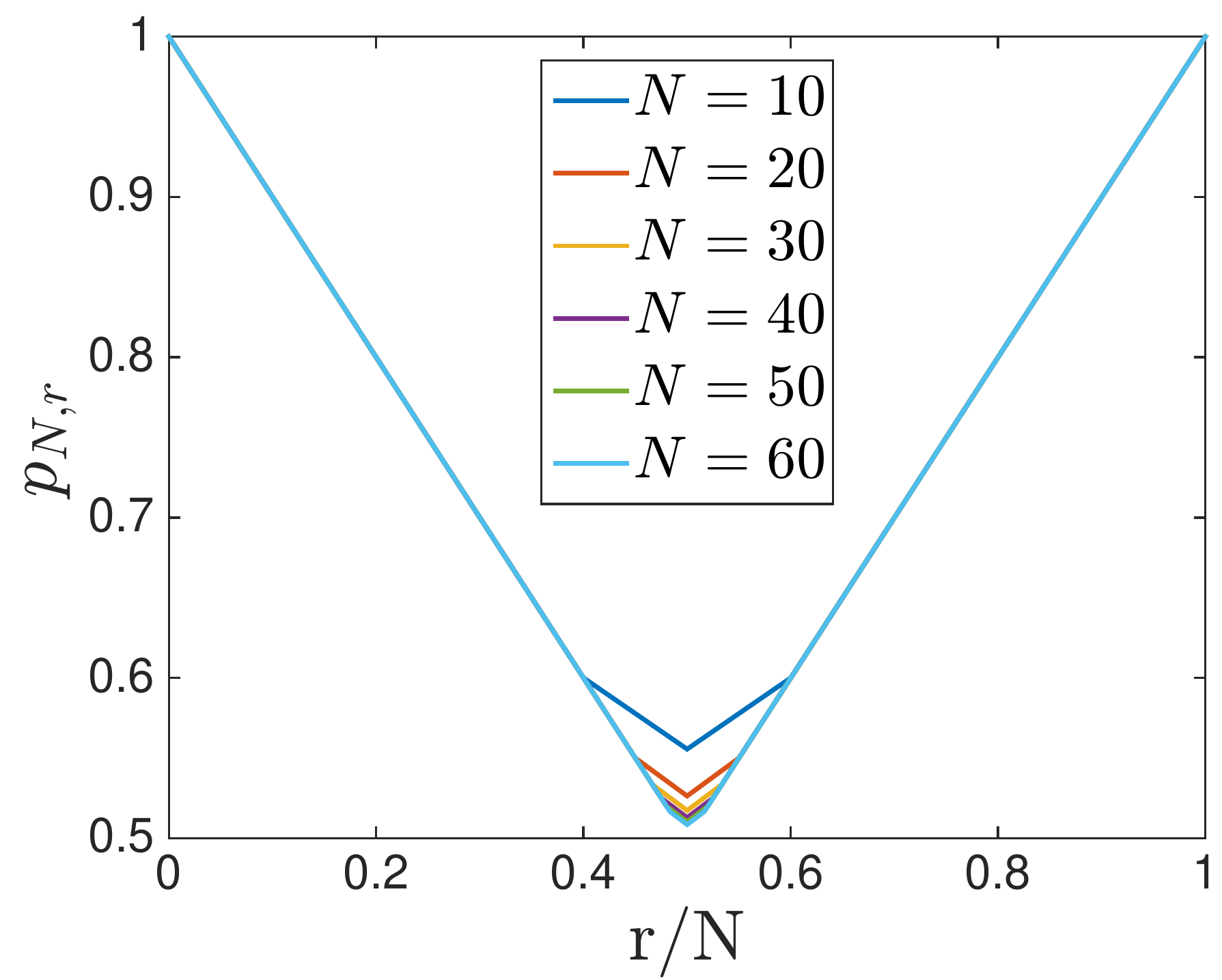}}
\subfloat[]{
\includegraphics[width=40mm,height=35mm]{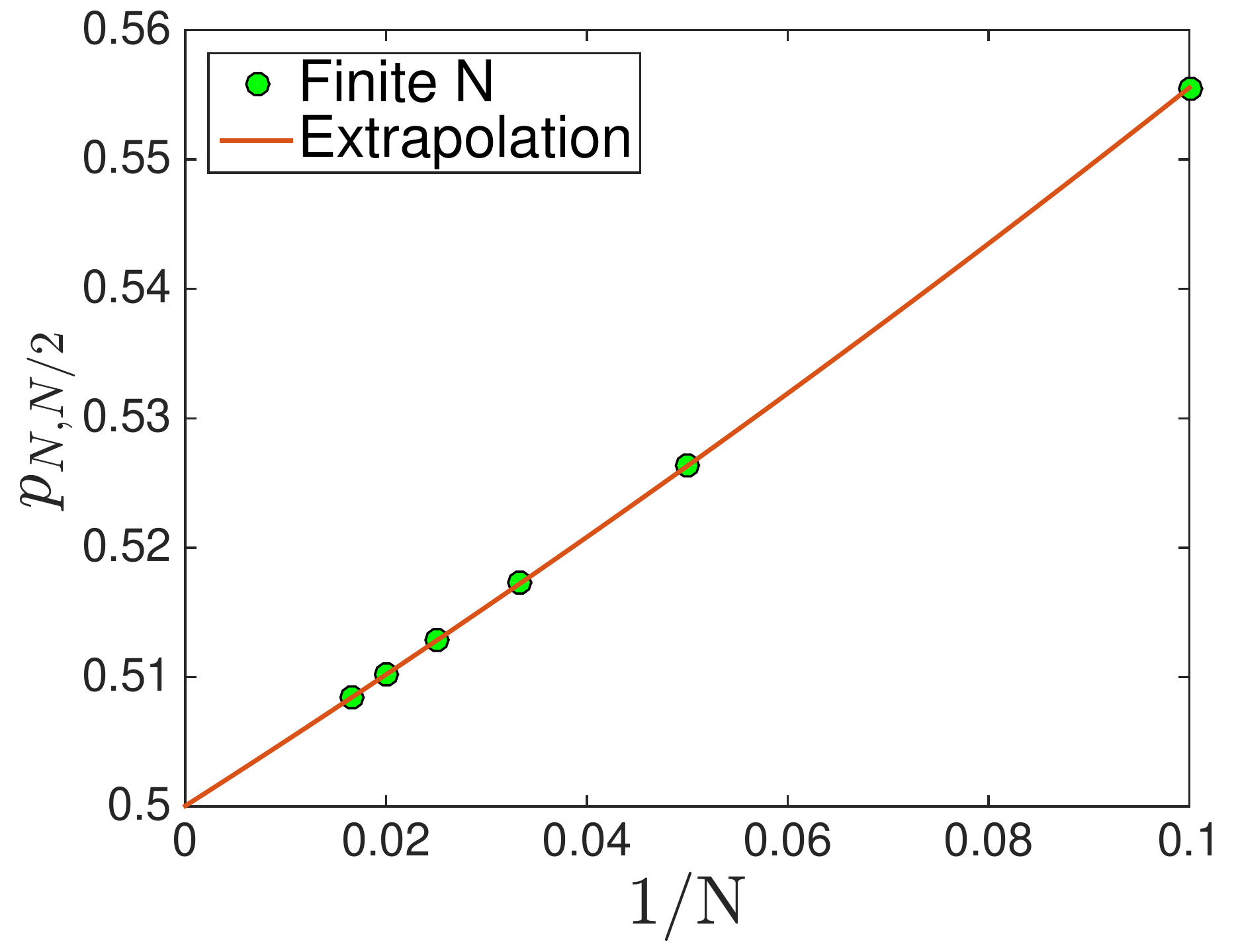}}
\caption{(a) The value of $p_{N,r}$ for different $N$ and $r$. The vertical axis is $p_{N,r}$,
and the horizontal axis is the $r/N$.
(b) The value of $p_{N,N/2}$ for twin Fock state $\ket{D_{N,N/2}}$ and its finite size extrapolation.}
\label{fig:maxvalue}
\end{figure}

We leave the detailed derivation of $p_{N,r}$ in Appendix B. Here we demonstrate the power of this observation by plotting the values of $p_{N,r}$ in Fig.~\ref{fig:maxvalue}(a) where it is shown up to $N=60$. Based on the finite $N$ scaling, it is expected that the $N=60$ line is very near the limit case of $N\rightarrow\infty$. This result also clearly shows that the twin Fock state $\ket{D_{N,N/2}}$ is the most stable one among all
Dicke states against arbitrary noise. We plot the value of $p_{N,N/2}$ for twin
twin Fock state in Fig.~\ref{fig:maxvalue}(b). From its finite $N$ scaling,
we infer that when $N\rightarrow\infty$, the limit value of $p_{N,N/2}$ will be at $1/2$.
That is, for any large system size $N$, a ratio of the population on twin Fock state $> 50\%$ almost ensures an entanglement depth $N$, for arbitrary noise.

The noise may be of a better form than an arbitrary noise in real experiments.
For instance, the white noise given by
$
\rho_{\text{noise}}=\frac{I}{2^N}.
$
One example is that for the twin Fock state $\ket{D_{N,N/2}}$ under white noise,
the entanglement depth of $\tilde{\rho}$ remains to be $N$ for $n_r>\sim \frac{1}{\sqrt{N}}$~\cite{ananth2015non}, which is in fact very close to a maximally mixed state. Another example is that the $N$-particle $W$ state $\ket{D_{N,1}}$, which is relatively unstable against white noise, is stable against all zero noise $\ket{0}^{\otimes N}$, where entanglement depth of $\sim 3000$ is reported in~\cite{mcconnell2015entanglement}.

The technique of measuring the occupation number $N_r$ of the Dicke state $\ket{D_{N,r}}$ is readily available in many
systems. For instance, the Mach-Zehnder interferometer in optical systems~\cite{demkowicz2015chapter,bachor2004guide}, and the one with similar principle
with atomic ensemble in double-well potentials~\cite{schumm2005matter,sebby2007preparing}.
For cold atomic ensembles with internal states, number occupation measurements are reported in~\cite{mcconnell2015entanglement} for the first few excited Dicke states, and for the vicinity of
twin Fock state are reported in~\cite{lucke2014detecting}.

In practice, the population is concentrated around some Dicke states with fluctuation (e.g. the distribution
of population around a twin Fock state as in ~\cite{lucke2014detecting}). It is then more realistic to model the state of the system as a mixture of Dicke states with noise, as given by
\begin{equation}
\label{eq:DickeM}
\tilde{\rho}=n_r\rho_{\text{Dicke}}+(1-n_r)\rho_{\text{noise}}.
\end{equation}
Here $\rho_{\text{Dicke}}$ is a mixture of some Dicke state
\begin{equation}
\label{eq:Dis}
\rho_{\text{Dicke}}=\sum_{r\in X} c_r\ket{D_{N,r}}\bra{D_{N,r}},
\end{equation}
with $X$ be a subset of integers, $c_r\geq 0$ and $\sum_r c_r=1$.

For preparing the desired Dicke state $\ket{D_{N,r_0}}$,
the $r$ appears in the summation distributed in Eq.~\eqref{eq:Dis} should be near the desired value $r_0$ (e.g. $r_0=N/2$ for twin Fock state). Using a similar method for obtaining the threshold value $p_{N,r}$,
we can also find a bound for a threshold value for $n_r$ in Eq.~\eqref{eq:DickeM}. The detailed analysis are included in Appendix B.

\section{Comparison with other entanglement detection methods}

Our results are in a sharp contrast with spin squeezing, which is only possible for a very small set of symmetric states for large system size $N$. More precisely, in order to detect entanglement of the state $\rho_N$ by the spin-squeezing criteria, the two-particle reduced density matrix (2-RDM) of $\rho_N$, denoted by $\rho_2$, must be entangled~\cite{wang2003spin}. However, if the state is symmetric, this is very unlikely
due to the quantum finite de Finetti's theorem~\cite{stormer1969symmetric,hudson1976locally,doherty02a,Renner2007,Christandl2007,harrow2013church}, which states that any $\rho_2$ of a symmetric $\rho_N$ is very close to a separable state, with the distance at most $\frac{1}{N}$ (see Appendix C for more details).

As an example, we discuss the tolerance of the Dicke states $\ket{D_{N,r}}$ with white noise $\rho_{\text{noise}}=I/2^N$. We examine
the separability of the corresponding $2$-RDM of $\tilde{\rho}$ to find the threshold value of $p'_{N,r}$, such that any $n_r>p'_{N,r}$ ensures that the $2$-RDM of $\tilde{\rho}$ is entangled.

We plot the value of $p'_{N,r}$ for Dicke states $\ket{D_{N,r}}$
in Fig.~\ref{fig:2RDM}(a) (see Appendix D for derivation of $p'_{N,r}$). We can see that when the number of particles $N$ increases,
the tolerance of noise decays quickly to near zero. This is indeed
in a sharp contrast with Fig.~\ref{fig:maxvalue}(a), where for many Dicke states that
the tolerance of noise does not decay much when the system size $N$ increases,
and approach a constant value in the limit $N\rightarrow\infty$.

For the twin Fock state $\ket{D_{N,N/2}}$ with white noise, we plot the threshold value of
 $p'_{N,N/2}$ vs. the particle number $N$, as given in
Fig.~\ref{fig:2RDM}(b).
In this case, the $2$-RDM for the noisy twin Fock state becomes
separable for $1-n_r>\sim \frac{1}{N}$ (meaning the population $p$ is almost $100\%$ for large $N$). This is in sharp contrast with the fact that the state $\tilde{\rho}$ is fully
entangled (i.e. has entanglement depth $N$) for $n_r>1/2$ for arbitrary noise and
 $n_r>\sim \frac{1}{\sqrt{N}}$ for white noise, as discussed above.

\begin{figure}[hbpt]
\centering
\subfloat[]{
\includegraphics[width=40mm,height=35mm]{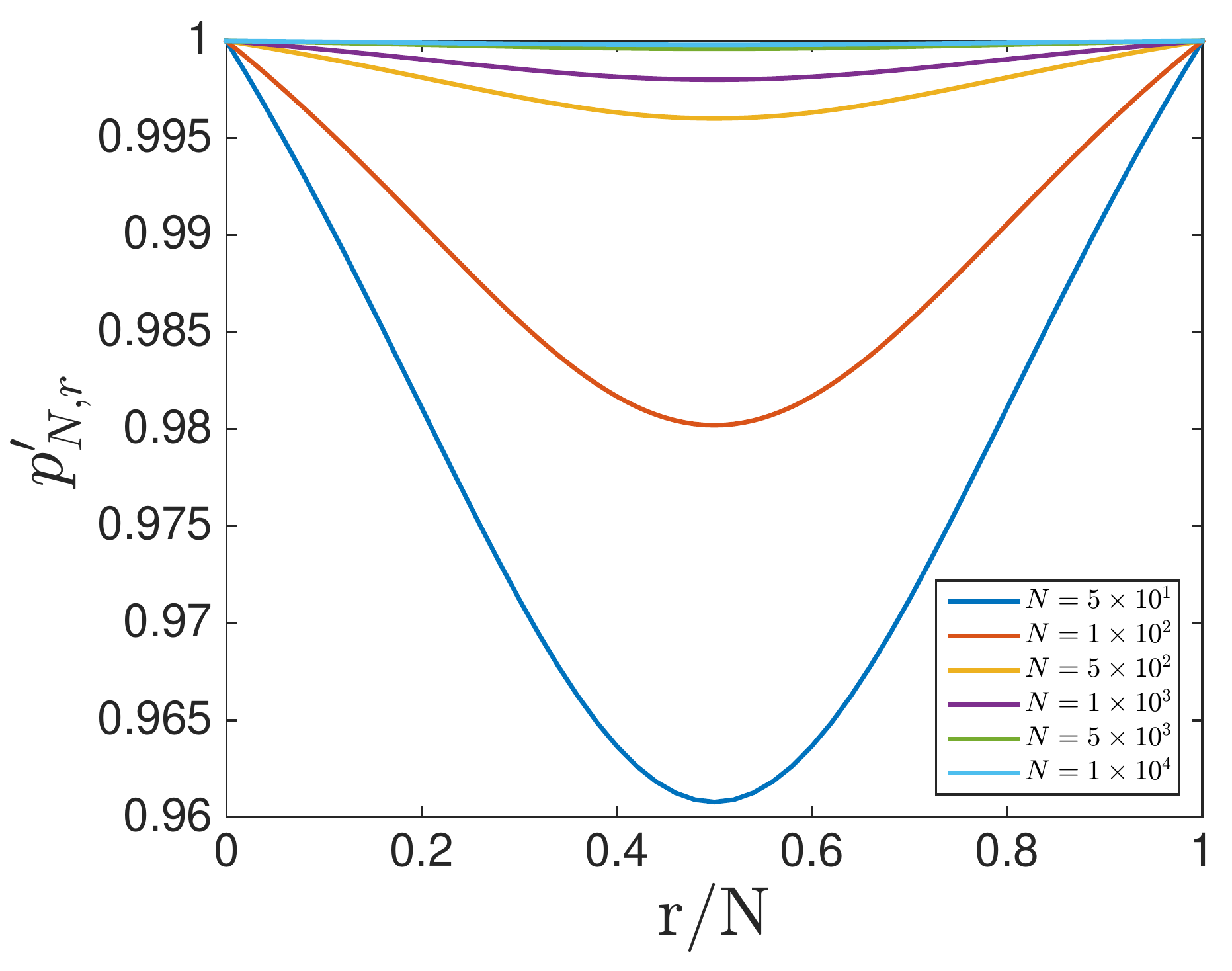}}
\subfloat[]{
\includegraphics[width=40mm,height=35mm]{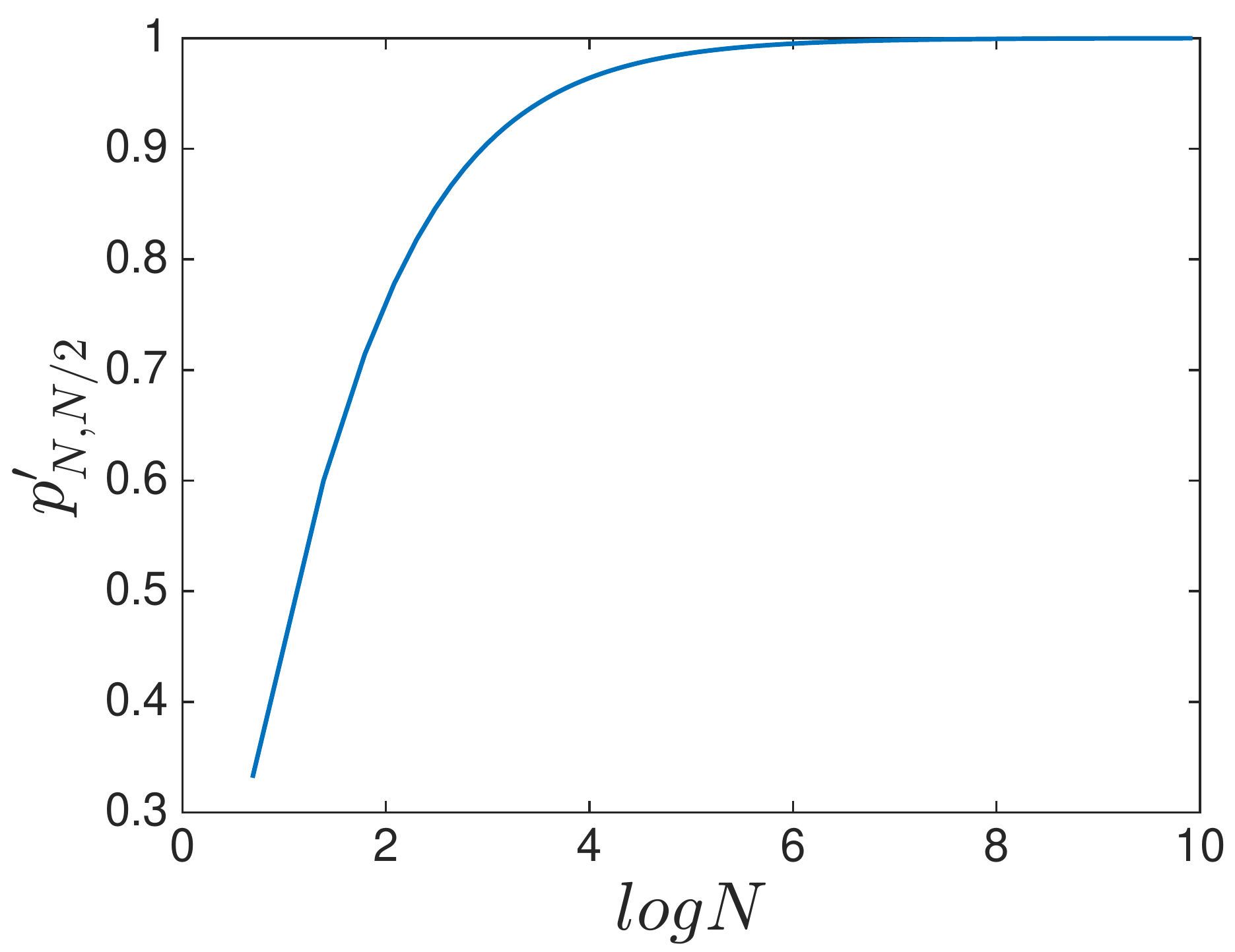}}
\caption{(a)$p'_{N,r}$ vs. $r$ for fixed $N$, for Dicke states $\ket{D_{N,r}}$.(b) $p'_{N,N/2}$ vs. $N$ for twin Fock state $\ket{D_{N,N/2}}$.}
\label{fig:2RDM}
\end{figure}

Another important measure for detecting entanglement is the quantum Fisher information $F(\rho_N)$, which is also an important measure for metrology properties of the $N$-particle state $\rho_N$~\cite{gross2012spin,toth2014quantum}. $F(\rho_N)\leq N$ for any separable $\rho_N$, so if $F(\rho_N)> N$ (to be useful for quantum metrology), then $\rho_N$ is necessarily entangled.
In general $F(\rho_N)$ is more sensitive to detect entanglement than spin squeezing. That is, $F(\rho_N)$ can detect some entanglement which can not be detected by spin squeezing~\cite{toth2014quantum}. Moreover, $F(\rho_N)$ is a lower bound of entanglement depth $\mathcal{D}(\rho_N)$, i.e. $F(\rho_N)\leq \mathcal{D}(\rho_N)N$~\cite{toth2012multipartite,hyllus2012fisher}.

Interestingly, it is recently shown that almost all symmetric states are useful for quantum metrology, while general entangled states are generally useless~\cite{oszmaniec2016random}. However, while both results state that symmetric states have some special entanglement properties, it is known that there is no dichotomy for $F(\rho_N)$, since there does exist entangled symmetric state (hence with entanglement depth $N$) that has $F(\rho_N)\leq N$~\cite{mcconnell2015entanglement}. This indicates that $F(\rho_N)$ is also not strong enough for detecting entanglement depth in general.

On the other hand, however, for entangled symmetric state, our dichotomy result gives $\mathcal{D}(\rho_N)=N$. In other words, for a symmetric state $\rho_N$, if it is detected as entangled by measuring either a spin squeezing parameter or $F(\rho_N)$, it is then fully entangled, i.e. $\mathcal{D}(\rho_N)=N$.

\section{Discussion} 

In this work, we have studied the entanglement depth for symmetric states. There is a dichotomy of entanglement depth for symmetric states: an $N$-particle symmetric state is either fully separable, or fully entangled -- i.e. the entanglement depth is either $1$ or $N$, an intrinsic property arising from permutation symmetry.  We show that this dichotomy property of entangled symmetric states is even stable under non-symmetric noise. We propose an experimentally accessible method to detect entanglement depth in atomic ensembles based on a bound on the particle number population of Dicke states, and demonstrate that the entanglement depth of some Dicke state, for example the twin Fock state, is very stable even under a large arbitrary noise.

Our results can be applied to infer entanglement depth in experimental situations where the state of the system
is near a symmetric state, such as the Bose-Einstein condensates. For instance,
our result shows that the twin Fock state $\ket{D_{N,N/2}}$ is very stable, where a population
of the twin Fock state slightly above $50\%$ will guarantee an entanglement depth
of the system size $N$. Since twin Fock state have been prepared in many experiments~\cite{kuzmich1998atomic,rodriguez2007generation,lucke2014detecting,smerzi2011twin}, our method shed light on characterizing entanglement properties of these practical systems.

\begin{acknowledgments}
{\bf Acknowledgments.} We thank Xinyu Luo, Zhaohui Wei, Ling-Na Wu, Zhi-Fang Xu, Meng Khoon Tey, Li You, Hui Zhai, and Duanlu Zhou for helpful discussions. We thank G{\'e}za T{\'o}th for bringing
Refs.~\cite{ichikawa2008exchange,wei2010exchange,toth2009entanglement} into our attention.
The work of JYC is supported by NSFC 11374176 and NSERC. NY and BZ are supported by NSERC. This research was supported in part by Perimeter Institute for Theoretical Physics. Research at Perimeter Institute is supported by the Government of Canada through Industry Canada and by the Province
of Ontario through the Ministry of Economic Development \& Innovation.
\end{acknowledgments}

\bibliography{depth}

\appendix

\section{Derivation of $p_{N,r}$}

We give the procedure for obtaining $p_{N,r}$ here. We start from the observation that
for any bipartite state with Schmidt decomposition $\ket{\psi}=\sum_{i}\sqrt{\lambda_i}\ket{i}\ket{i}$ and $\lambda_0\geq \lambda_1\geq,\cdots$, its overlap with product state is no more than $\lambda_0$, where the overlap of two pure state $\ket{\alpha},\ket{\beta}$ is defined as $|\ip{\alpha}{\beta}|^2$.

To do so, let integers $m_0,m_1\geq 1$ satisfy $m_0+m_1=N$, one can verify the following Schmidt decomposition,
\begin{eqnarray*}
\ket{D_{N,r}}&=&\sum_{j=\max\{0,r-m_1\}}^{\min\{r,m_0\}}\sqrt{\frac{C_{m_0}^{j}C_{m_1}^{r-j}}{C_N^{r}}}\ket{D_{m_0,j}}\ket{D_{m_1,r-j}}.
\end{eqnarray*}

Suppose a bi-separable state $\rho_N=\sum_{i} p_i \op{\psi_i}{\psi_i}$ with product $\ket{\psi_i}$. According to the above observation, we know that $|\ip{\psi_i}{D_{N,r}}|^2\leq p_{N,r}$, where $p_{N,r}=\max\{\frac{C_{m_0}^{j}C_{m_1}^{r-j}}{C_N^{r}}\}$. Therefore, $\langle D_{N,r}|\rho_N\ket{D_{N,r}}\leq \sum_i p_i p_{N,r}=p_{N,r}$.

Now consider a mixture of Dicke states
\begin{eqnarray*}
\rho_{\text{Dicke}}=\sum_{r\in X} c_r\ket{D_{N,r}}\bra{D_{N,r}},
\end{eqnarray*}
with $c_r\geq 0$ and $\sum_r c_r=1$.
Define $q_X:=\max_{\ket{\psi}} \langle\psi|P_X|\psi\rangle$ with $\ket{\psi}$ ranging over all bi-product state among all bipartition of $N$-qubit, and $P_X:=\sum_{r\in X} \ket{D_{N,r}}\bra{D_{N,r}}$.
Then for a bi-separable state $\rho_N=\sum_{i} p_i \op{\psi_i}{\psi_i}$ with product $\ket{\psi_i}$,
we have
\begin{eqnarray*}
&&\tr(\rho_NP_X)\\&=&\tr(\sum_{i} p_i (\op{\psi_i}{\psi_i})P_X)\\
&\leq&\max_{\ket{\psi}} \langle\psi|P_X|\psi\rangle\\
&=&q_X.
\end{eqnarray*}
That is, if $p>q_X$ for arbitrary noise, the state $\tilde{\rho}$ is fully entangled, i.e. has entanglement depth
$N$. And the distribution
of $c_r$ can be obtained by the population on each Dicke state $\ket{D_{N,r}}$.
In other words, as long as one ensures that the number of total population on the two modes
of interest are larger than $Nq_X$, then the state of the system is fully entangled.

\section{Entanglement depth and spin squeezing}

It has been known that spin-squeezing and its generalizations also provide methods to detect entanglement depth~\cite{sorensen2001entanglement,lucke2014detecting,gross2012spin,toth2014quantum,
duan2011entanglement,korbicz2006generalized,ma2011quantum,vitagliano2014spin,korbicz2005spin}.
However, we will clarify that this method is in general ineffective for symmetric states due to the finite quatnum de Fenitti's theorem~\cite{stormer1969symmetric,hudson1976locally,doherty02a,Renner2007,Christandl2007,harrow2013church}.

Consider the collective spin operator
\begin{equation}
\vec{J}=\sum_{i=1}^N\vec{S}_i,
\end{equation}
where each spin operator $\vec{S}_i$ acts on the $i$th particle.
Denote the component of $\vec{J}$ by $(J_x,J_y,J_z)$, and
their variations by $(\Delta J_x,\Delta J_y, \Delta J_z)$.
The idea of detecting entanglement depth by spin squeezing is to
compute the value of some inequalities of $(J_x,J_y,J_z)$ and their variations,
which can reveal the entanglement depth of the system.

For any $N$-particle state $\rho_N$, notice that
\begin{equation}
\Delta J_x = \tr(J_x^2\rho_N)- (\tr(J_x\rho_N))^2,
\end{equation}
where $J_x^2$ involves only two-particles interactions, therefore both $J_x$ and its variation
(similar for $J_y$ and $J_z$) can be calculated from the two-particle reduced density matrices of
$\rho_N$. In order to detect the entanglement of $\rho_N$ the
two-particle reduced density matrices of $\rho_N$ must be entangled.
In other words, {\it any  $N$-particle spin-squeezed state $\rho_N$ is pairwise entangled} (which is observed for the standard spin squeezing in~\cite{wang2003spin}).

However, for symmetric states, unfortunately, their $k$-particle reduced density matrices ($k$-RDMs) are almost separable (i.e. not entangled) for large $N$. This fact is given by the celebrated
finite quantum de Finetti's theorem~\cite{stormer1969symmetric,hudson1976locally,doherty02a,Renner2007,Christandl2007,harrow2013church}. In particular for $k=2$, the $2$-RDM $\rho_2$ of any symmetric $N$-particle state $\rho_N$ satisfies
\begin{equation}
|\rho_2-{\rho}^s_2|<\frac{2d}{N},
\end{equation}
where ${\rho}^s_2$ is an arbitrary two-particle separable state. Here $d$ is the dimension of
the single-particle space. In other words, $2$-RDM $\rho_2$ is very close to separable states
up to a distance scales as $\sim \frac{1}{N}$. The larger $N$, the less likely that $\rho_2$ can be
entangled (hence $\rho_N$ can be spin-squeezed).

\section{Derivation of $p'_{N,r}$}
For the state $\tilde{\rho}$ with $\rho_{\text{noise}}=I/2^N$, it is easy to see the 2-RDM $\rho_2$ takes the following form:
\begin{equation}\nonumber
\rho_2=\frac{n_r}{N(N-1)}\cdot
\begin{pmatrix}
    \rho_{00} & 0 & 0 & 0\\
    0 & \rho_{01} & \rho_{01} & 0\\
    0 & \rho_{01} & \rho_{01} & 0\\
    0 & 0 & 0 & \rho_{11}
\end{pmatrix}
+\frac{1-n_r}{4}\cdot I_4,
\end{equation}
where
\begin{eqnarray*}
\rho_{00}&=&(N-r)^2-(N-r), \\
\rho_{11}&=&r^2-r, \\
\rho_{01}&=&(N-r)r
\end{eqnarray*}
 and $I_4$ is $4\times 4$ identity matrix. The partial transpose of $\rho_2$ is given by:
\begin{equation}\nonumber
\rho_2^{T_B}=\frac{n_r}{N(N-1)}\cdot
\begin{pmatrix}
    \rho_{00} & 0 & 0 & \rho_{01} \\
    0 & \rho_{01} & 0 & 0 \\
    0 & 0 & \rho_{01} & 0 \\
    \rho_{01} & 0 & 0 & \rho_{11}
\end{pmatrix}
+\frac{1-n_r}{4}\cdot I_4.
\end{equation}

$p'_{N,r}$ is given by the condition that the smallest eigenvalue of $\rho_2^{T_B}$ equals to zero \cite{horodecki1996separability}. Explicitly, we have:
\begin{equation}
p'_{N,r}=\frac{N(N-1)}{N(N-1)+2c},
\end{equation}
where
\begin{eqnarray*}
&&c=-[(N-r)^2+r^2-N]\nonumber\\
&&+\sqrt{4(N-r)^2r^2+(N-2r)^2(N-1)^2}.
\end{eqnarray*}

For the case, $r=N/2$, we find that
\begin{equation}
p'_{N,N/2}=\frac{N-1}{N+1}=1-2/(N+1).
\end{equation}

\end{document}